\documentclass[apj]{emulateapj}
\usepackage{subfigure}
\usepackage{graphics}

\def\spose#1{\hbox to 0pt{#1\hss}}
\def\simlt{\mathrel{\spose{\lower 3pt\hbox{$\mathchar"218$}}
    \raise 2.0pt\hbox{$\mathchar"13C$}}}
\def\simgt{\mathrel{\spose{\lower 3pt\hbox{$\mathchar"218$}}
    \raise 2.0pt\hbox{$\mathchar"13E$}}}

\newcommand{\oiii}{\mbox{[\ion{O}{3}]} $\,$}
\newcommand{\oiiiw}{\mbox{[\ion{O}{3}] $\lambda$5007} $\,$}
\newcommand{\oiiiwn}{\mbox{[\ion{O}{3}] $\lambda$5007}}

\newcommand{\hb}{\mbox{H$\beta$} $\,$}
\newcommand{\hbn}{\mbox{H$\beta$}}
\newcommand{\ha}{\mbox{H$\alpha$} $\,$}
\newcommand{\han}{\mbox{H$\alpha$}}

\newcommand{\nii}{\mbox{[\ion{N}{2}] $\lambda$6584} $\,$}

\slugcomment{\it Accepted for Publication in ApJ}
\shortauthors{Comerford \& Greene}
\shorttitle{Offset AGNs as Tracers of Galaxy Mergers and Supermassive Black Hole Growth}

\begin{document}

\title{Offset Active Galactic Nuclei as Tracers of Galaxy Mergers \\ and Supermassive Black Hole Growth}

\author{Julia M. Comerford\altaffilmark{1} and Jenny E. Greene\altaffilmark{2}}

\affil{$^1$Department of Astrophysical and Planetary Sciences, University of Colorado, Boulder, CO 80309, USA}
\affil{$^2$Department of Astrophysical Sciences, Princeton University, Princeton, NJ 08544, USA}

\begin{abstract}
Offset active galactic nuclei (AGNs) are AGNs that are in ongoing galaxy mergers, which produce kinematic offsets in the AGNs relative to their host galaxies. Offset AGNs are also close relatives of dual AGNs.  We conduct a systematic search for offset AGNs in the Sloan Digital Sky Survey, by selecting AGN emission lines that exhibit statistically significant line-of-sight velocity offsets relative to systemic.  From a parent sample of 18314 Type 2 AGNs at $z<0.21$, we identify 351 offset AGN candidates with velocity offsets of 50 km s$^{-1} < |\Delta v| < 410$ km s$^{-1}$.  When we account for projection effects in the observed velocities, we estimate that $4\%$ -- $8\%$ of AGNs are offset AGNs.  We designed our selection criteria to bypass velocity offsets produced by rotating gas disks, AGN outflows, and gravitational recoil of supermassive black holes, but follow-up observations are still required to confirm our candidates as offset AGNs.  We find that the fraction of AGNs that are offset candidates increases with AGN bolometric luminosity, from $0.7\%$ to $6\%$ over the luminosity range $43 < \log(L_\mathrm{bol})$ [erg s$^{-1}$] $< 46$.  If these candidates are shown to be bona fide offset AGNs, then this would be direct observational evidence that galaxy mergers preferentially trigger high-luminosity AGNs.  Finally, we find that the fraction of AGNs that are offset AGN candidates increases from $1.9\%$ at $z=0.1$ to $32\%$ at $z=0.7$, in step with the growth in the galaxy merger fraction over the same redshift range.
\end{abstract}  


\section{Introduction}
\label{intro}

A merger between two galaxies, each with its own central supermassive black hole (SMBH), brings the two SMBHs to the center of the resultant merger-remnant galaxy.  The pair is known as dual SMBHs when the black holes are separated by kiloparsec (kpc) scales, before the pair evolves into a gravitationally-bound SMBH binary and ultimately coalesces.

Since galaxy mergers can trigger gas inflows that fuel active galactic nuclei (AGNs; \citealt{SU98.1,CA01.2,TR12.1}) and the AGN fraction increases as the SMBH separation decreases from $\sim80$ kpc to 5 kpc \citep{EL11.1,KO12.1}, some of these dual SMBHs should be active.  When one or both SMBHs power AGNs, the systems are known as offset AGNs and dual AGNs, respectively.  These offset and dual AGNs are valuable for studies of galaxy evolution, since they are direct observational tracers of SMBH mass growth via gas accretion during mergers.

Dual AGNs have been popular targets of recent study (e.g., \citealt{CO09.3,BA12.1,KO12.1,TE12.1,VA12.1,BL13.1,IM14.1,WO14.1}), with the first systematic searches beginning in the last few years.   Most of these searches focus on AGN spectra with double-peaked narrow emission lines (e.g., \citealt{CO09.1,WA09.1,XU09.1,LI10.1,SM10.1,GE12.1,BA13.1,CO13.1}), which can be produced by the relative motions of two narrow-line regions (NLRs) accompanying two AGNs moving in the host galaxy potential.  

Double-peaked narrow AGN emission lines can can also be caused by disk rotation and the NLR structure of biconical AGN outflows (e.g., \citealt{HE81.1,CR00.1,VE01.1,WH04.1,DA05.1,CR10.1,FI11.1}).  To determine the true nature of double-peaked AGNs there have been many follow-up observations, including optical spectroscopy, near infrared imaging, radio observations, {\it Hubble Space Telescope} imaging, and {\it Chandra} observations \citep{LI10.2, CO11.2, FU11.1,FU11.3,GR11.1,MC11.1,RO11.1,SH11.1,TI11.1,CO12.1,FU12.1,GR12.2,LI13.1}.  These follow-up observations have led to several confirmations of double-peaked emission lines that are produced by dual AGNs \citep{FU11.3,LI13.1}.

In contrast, offset AGN candidates have so far received little attention despite their potential for yielding many more dual SMBH discoveries.  Study of offset AGNs will also open the door to comparisons between offset AGN and dual AGN populations, which will uncover the details of AGN fueling during mergers and differing circumstances for fueling of one versus both SMBHs in a merger.  

Offset AGNs have been observed, as in the example of the $z=0.0271$ disturbed disk galaxy NGC 3341 \citep{BA08.1,BI13.1}.  This offset AGN was a serendipitous discovery, and it is a Seyfert 2 at a projected separation of $9\farcs5$ (5.2 kpc) from the nucleus of the host galaxy.  The offset AGN also has a line-of-sight velocity that is blueshifted by 200 km s$^{-1}$ relative to the host galaxy nucleus.  Several studies of AGNs in galaxy pairs, where pairs are defined by maximum transverse separations varying from 30 to 80 kpc and maximum line-of-sight velocity separations varying from 200 to 500 km s$^{-1}$, may have also detected offset AGNs, although they are not discussed specifically \citep{AL07.1,WO07.2,RO09.1,KO10.1,EL11.1}.

We conduct a systematic search for offset AGNs in the Sloan Digital Sky Survey (SDSS; \citealt{YO00.2}) via the spectral signature of AGN emission lines that are offset in line-of-sight velocity from systemic, as in NGC 3341.  This approach is analogous to the technique of searching for dual AGNs via double-peaked narrow AGN emission lines.  Similar to the case for double-peaked AGN emission lines, single-peaked AGN emission lines with velocity offsets can be produced by disk rotation, AGN outflows, recoiling SMBHs, dust obscuration, or dual SMBHs.  Consequently, the velocity-offset AGNs we find here are candidates for offset AGNs, but confirmation of their true natures requires additional follow-up observations. 
 
\begin{deluxetable*}{lllll}
\tabletypesize{\scriptsize}
\tablewidth{0pt}
\tablecolumns{5}
\tablecaption{Summary of Offset AGN Candidates} 
\tablehead{
\colhead{SDSS Designation} &
\colhead{Host Galaxy Redshift} & 
\colhead{$\Delta v_\mathrm{Balmer}$ (km s$^{-1}$)} &
\colhead{$\Delta v_\mathrm{forbidden}$ (km s$^{-1}$)} &
\colhead{$\Delta v_\mathrm{weighted}$ (km s$^{-1}$)} 
}
\startdata
SDSS J001828.09$-$003412.3 & $0.069291 \pm 0.000017$ & $-64.4 \pm 14.7$ & $-70.0 \pm 14.7$ & $-67.2 \pm 10.4$ \\
SDSS J002312.35+003956.2 & $0.072648 \pm 0.000020$ & \phn \phd $47.9 \pm 14.8$ & \phn \phd $54.8 \pm 14.7$ & \phn \phd $51.4 \pm 10.4$ \\
SDSS J003908.37$-$105833.0 & $0.065406 \pm 0.000020$ & $-72.6 \pm 14.8$ & $-52.1 \pm 14.8$ & $-62.4 \pm 10.5$ \\
SDSS J003948.38$-$090834.5 & $0.037475 \pm 0.000023$ & \phn \phd $52.6 \pm 14.7$ & \phn \phd $49.7 \pm 14.7$ & \phn \phd $51.1 \pm 10.4$ 
\enddata
\tablecomments{(This table is available in its entirety in a machine-readable form in the online journal.  A portion is shown here for guidance regarding its form and content.)}
\label{tbl-1}
\end{deluxetable*}

This work is the third systematic search for offset AGNs.  The first two searches identified offset AGN candidates via velocity-offset AGN emission lines in the DEEP2 Galaxy Redshift Survey \citep{CO09.1} and in the AGN and Galaxy Evolution Survey (AGES; \citealt{CO13.1}).  These searches uncovered 30 offset AGN candidates at a mean redshift $\bar{z}=0.7$ and 5 offset AGN candidates at $\bar{z}=0.25$, respectively.  The search for offset AGN candidates in SDSS has the potential to yield many more candidates, given the much larger size of the SDSS spectroscopic catalog compared to those of DEEP2 and AGES.  Furthermore, the search through SDSS ($\bar{z}=0.1$) will fill in a population of offset AGN candidates at low redshifts, which will enable follow-up observations to resolve offset AGNs with $<$ kpc projected offsets. 

We note that while an offset AGN is the case of dual SMBHs where one SMBH is active and the other is quiescent, our search here focuses by necessity on {\it detectable} offset AGNs.  Detectable offset AGNs are dual SMBH systems where one SMBH is an AGN and the other SMBH is not detected as an AGN, either because it is a quiescent SMBH or an AGN that is obscured by dust or confused with star formation. 

We assume a Hubble constant $H_0 =70$ km s$^{-1}$ Mpc$^{-1}$, $\Omega_m=0.3$, and $\Omega_\Lambda=0.7$ throughout, and all distances are given in physical (not comoving) units.

\section{Selecting Offset AGN Candidates}
\label{selection}

We begin with the SDSS DR7 catalog of $z<0.21$ objects identified as galaxies by the SDSS pipeline \citep{AB09.1} and the OSSY catalog \citep{OH11.1} of velocity dispersion, line position, and flux measurements for SDSS DR7 spectra.  OSSY uses codes for penalized pixel-fitting ({\tt pPXF}) and gas and absorption line fitting ({\tt gandalf}; \citealt{SA06.1}) to 
simultaneously fit an entire spectrum using stellar templates for the stellar kinematics and Gaussian templates for the emission components.  This process returns high quality measurements of wavelengths, fluxes, and widths of absorption and emission features for each SDSS spectrum.  OSSY defines the quality of their fits to the spectra using the level of the formal uncertainties in the flux densities, which is the statistical noise, and the level of fluctuations in the fit residuals, which is the residual noise. 

From the catalog of $z<0.21$ galaxies in SDSS DR7, we select the 20098 spectra identified as Type 2 AGNs in \cite{BR04.1}, which requires that the \oiiiwn, \hbn, \han, and \nii lines have signal-to-noise ratios greater than 3 and that the emission line flux ratios lie above the theoretically derived boundary between composite systems and AGNs \citep{KE01.2}.  

We restrict the sample to those AGNs with robust fits to the absorption and emission line systems in the SDSS spectra.  To do this, we examine the plot of residual-noise-to-statistical-noise ratio against median signal-to-statistical-noise ratio for the OSSY catalog.  We define quality fits as those that deviate by less than $3\sigma$ from the median line in this plot (see \citealt{OH11.1}). This criterion reduces the sample to 18314 AGNs, which we define to be our ``parent sample" of AGNs from which we will identify offset AGN candidates.   

Our purpose is to select the AGN spectra with kinematic signatures of offset AGNs.  Specifically, we search for line-of-sight velocity offsets of the AGN-fueled emission lines relative to the stellar absorption features, because such velocity offsets are an expected consequence of the bulk motion of an AGN brought into a merger-remnant galaxy.  Since other effects such as AGN outflows, disk rotation, gravitational recoil of SMBHs, and dust obscuration are well known to produce velocity offsets in AGN emission lines, we carefully construct criteria that will select {\it for} offset AGNs and {\it against} these other kinematic effects.  

Our selection criteria for offset AGNs are as follows.  In the OSSY catalog, all forbidden lines are forced to have the same kinematics, while all the Balmer lines are fit with a separate kinematical model (e.g., \citealt{TR04.2}).  We use the velocities of the forbidden lines, Balmer lines, and stellar absorption features measured in OSSY to derive line-of-sight velocity offsets of the forbidden and Balmer lines relative to the stars. 

First, we require that the line-of-sight velocity offsets of the forbidden lines and of the Balmer lines agree to $1\sigma$.  This criterion aids in separating the offset-AGN candidates (where all emission lines should have consistent velocity offsets, due to the bulk motion of the AGN) from the AGN outflows (where there may be a stratified velocity structure, with different velocity offsets in the Balmer and forbidden lines; e.g., \citealt{ZA02.1}) and the gravitationally recoiling SMBHs (where velocity offsets are seen in the broad emission lines; e.g., \citealt{JU13.1}). We measure the line-of-sight velocity offsets relative to the measured redshift of the stellar absorption lines, which we take to be systemic.  In addition to the uncertainties on the line velocities reported in OSSY, we also add a systematic uncertainty based on the variations in multiple SDSS observations of the same system.  In the 1289 AGNs in our sample that were observed more than once with SDSS, we find a mean redshift difference of 10 km s$^{-1}$ between two epochs of observations.  Consequently, we add a 10 km s$^{-1}$ systematic uncertainty to each velocity.  We find that 15173 AGNs have forbidden lines and Balmer lines whose line-of-sight velocity offsets are consistent to $1\sigma$.

Second, we select the AGNs with line-of-sight velocity offsets that are greater than $3\sigma$ in significance.  This criterion is designed to select against stationary AGN and SDSS spectra that may have been taken with miscentered fibers (which could result in rotating gas producing a slight redshifted or blueshifted velocity offset in the emission lines), but as a side effect it also removes bona fide offset AGNs that have small line-of-sight velocity offsets due to projection effects (we account for this selection effect in Section~\ref{small}).  After this cut 544 AGNs remain, and we define them as the ``velocity-shifted sample" of AGNs.

Third, we require that the AGN emission line profiles are symmetric.  The goal here is to select against AGN outflows, whose commonly asymmetric line profiles have been well documented (e.g., \citealt{HE81.1,WH85.1,CR00.2,TA01.1,ZA02.1,GR05.1,DA06.1,KO07.2,WA11.1}).  To do this, we fit the spectra with {\tt gandalf} to obtain the continuum-subtracted emission line profiles.  Then, we borrow the approach used in studies of high redshift galaxies to distinguish asymmetric Ly$\alpha$ emission lines at high redshift from the symmetric emission lines, such as H$\alpha$, of lower redshift interlopers (e.g., \citealt{DI07.1}).  Following \cite{KA06.1}, we define the weighted skewness as the statistical skewness of a line's continuum-subtracted flux, over the wavelength range where the continuum-subtracted flux values are greater than $10\%$ of the line's peak continuum-subtracted flux value.  Since the blended \mbox{[\ion{N}{2}]} and \ha lines prevent accurate measurements of their skewnesses, we focus on the \hb and \oiii lines.   We require weighted skewnesses less than 0.5 \citep{BU79.1} in the continuum-subtracted \hb and \oiii emission lines, and these criteria are met by 365 AGNs.

\begin{figure}
\begin{center}
\includegraphics[width=8.5cm]{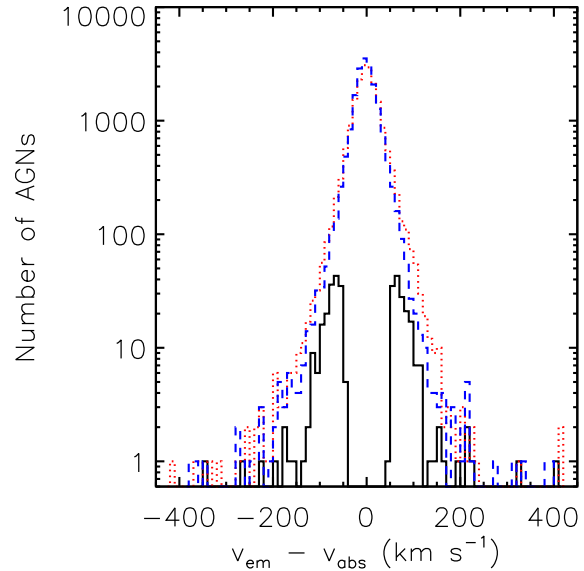}
\end{center}
\caption{Histograms of line-of-sight velocity offsets in the SDSS $z<0.21$ sample of AGNs.  For the parent sample of 18314 AGNs, we show the velocity offsets of the forbidden emission lines (red dotted) and the Balmer emission lines (blue dashed) relative to the stellar absorption features. The black histogram illustrates the weighted velocity offsets of the 351 offset AGN candidates.}
\label{fig:vhist}
\end{figure}

\begin{figure*}
		\hspace*{.4in}
		\includegraphics[width=.6\linewidth]{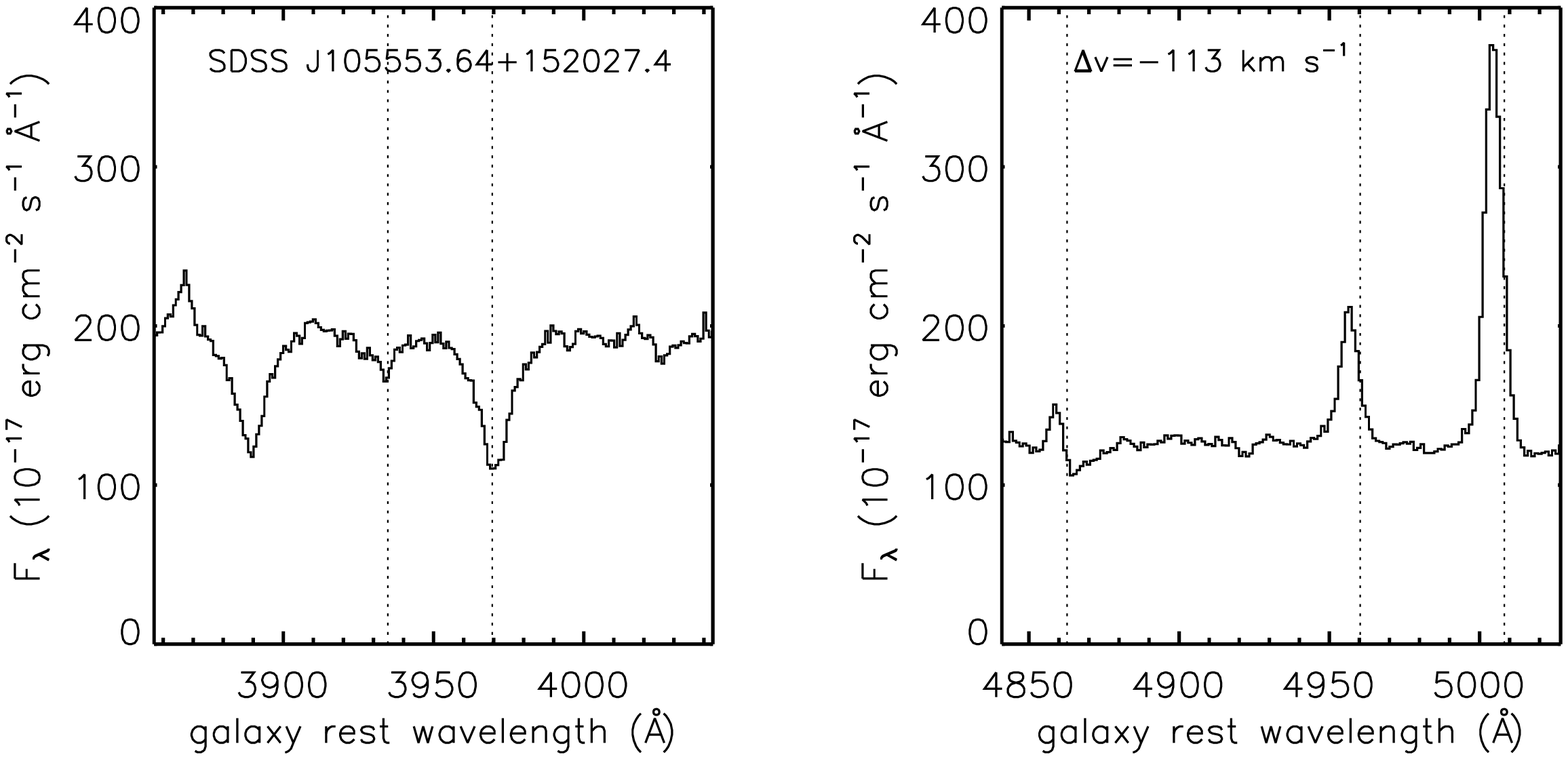}
		\includegraphics[width=.275\linewidth]{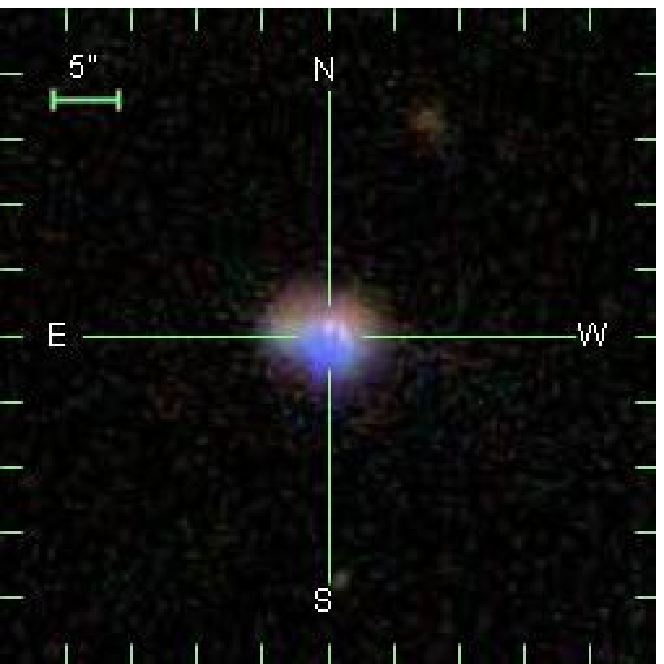} \\
		\hspace*{.4in}
		\includegraphics[width=.6\linewidth]{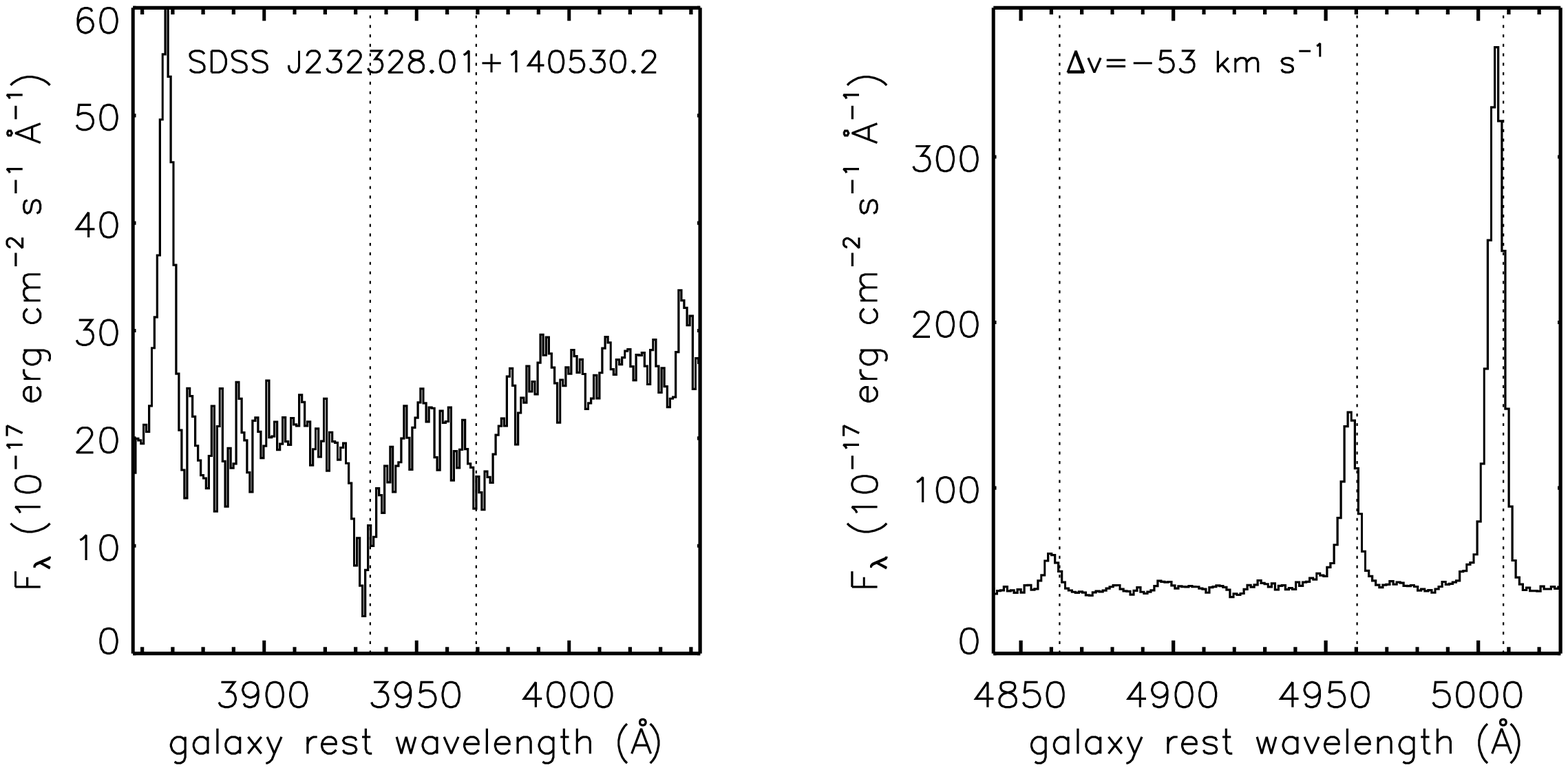}
		\includegraphics[width=.275\linewidth]{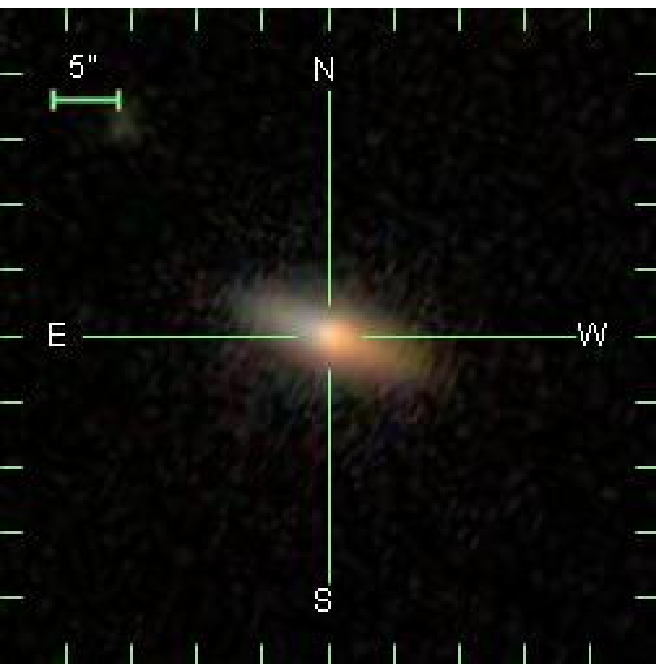} \\
		\hspace*{.4in}
		\includegraphics[width=.6\linewidth]{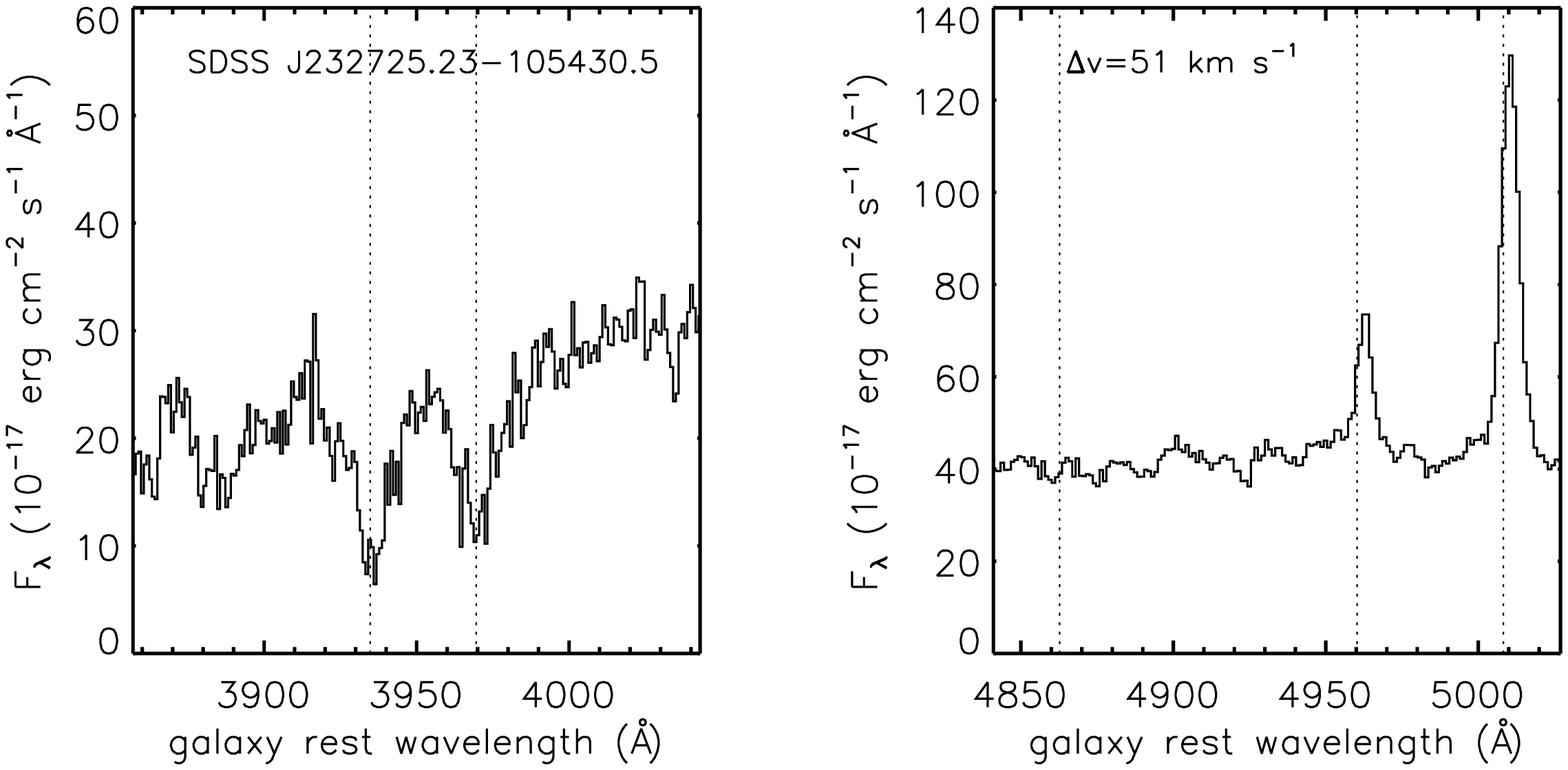}
		\includegraphics[width=.275\linewidth]{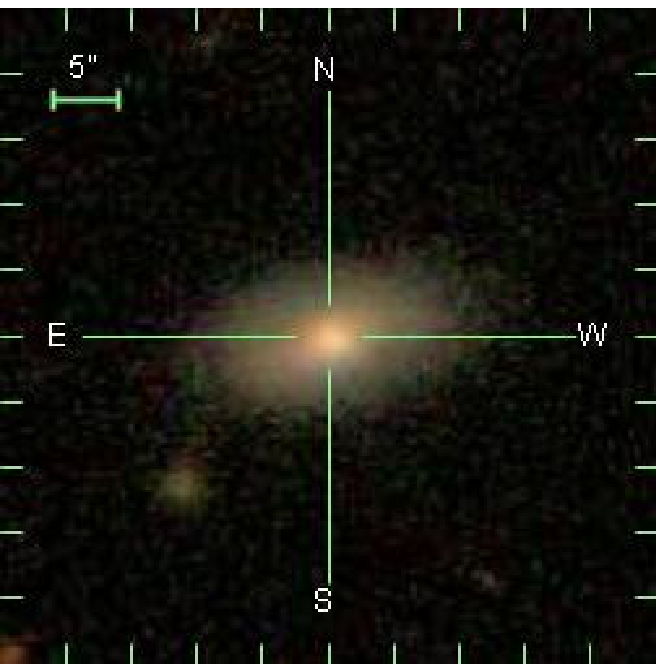} \\
		\hspace*{.4in}
		\includegraphics[width=.6\linewidth]{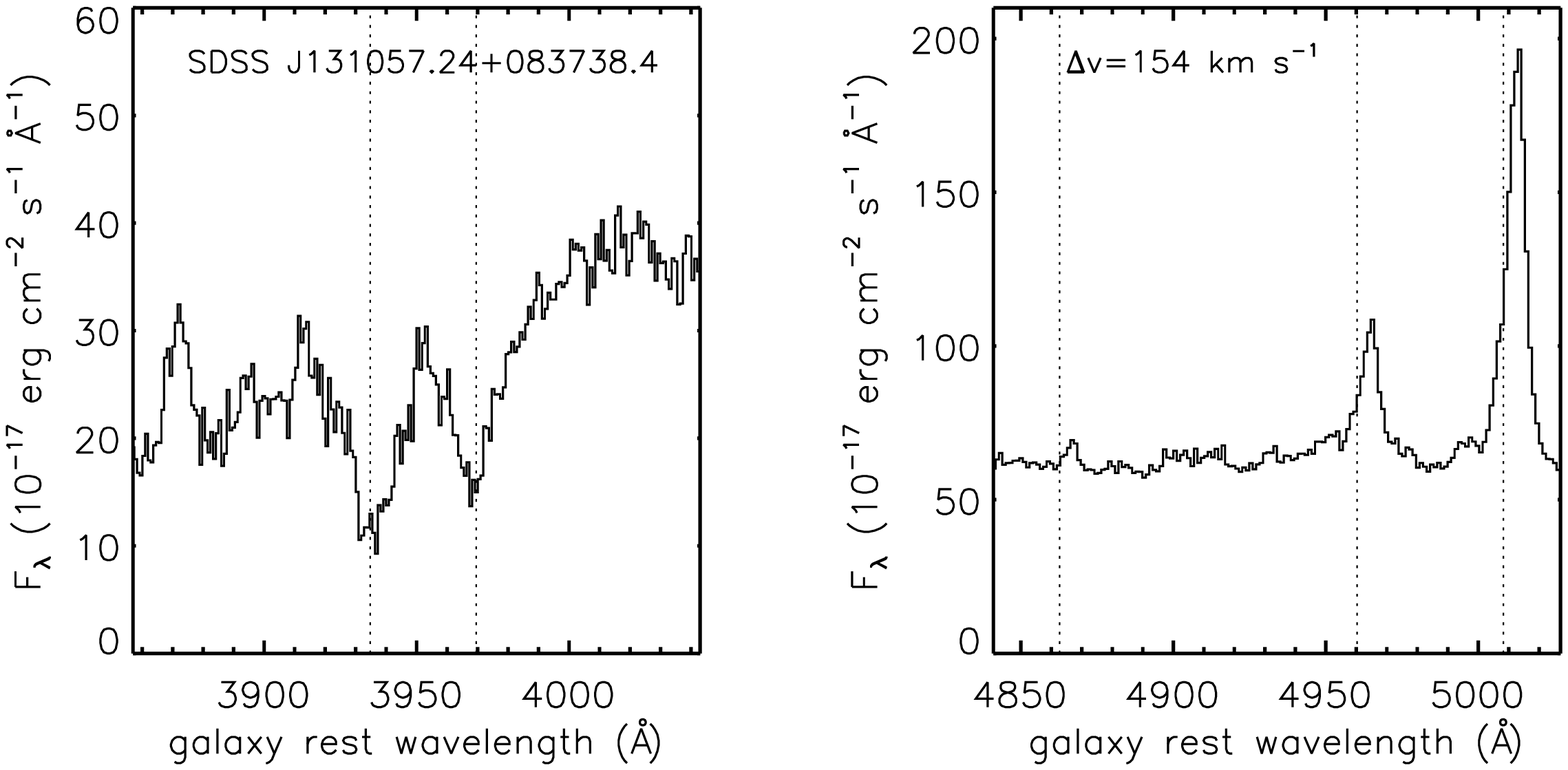}
		\includegraphics[width=.275\linewidth]{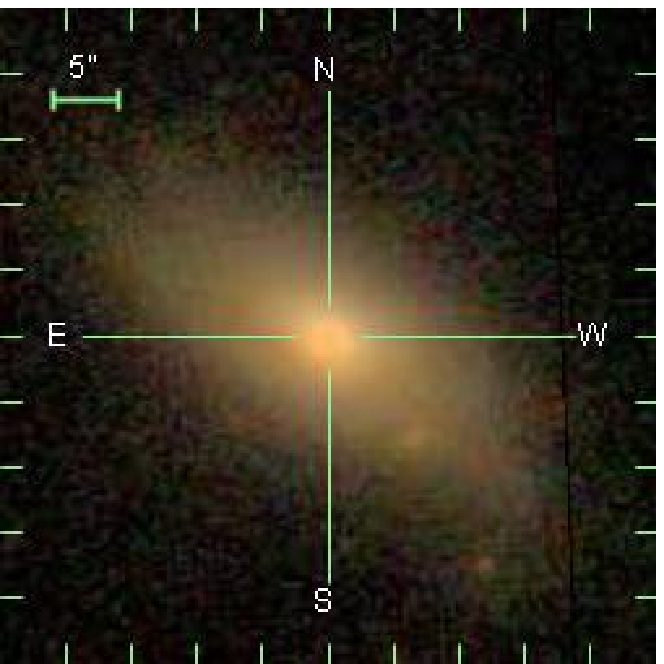}		
\caption{\scriptsize{SDSS spectra and imaging of four example offset AGN candidates.  Left: the SDSS spectra are plotted in the restframe of each galaxy, based on the redshift of the stellar absorption features (such as Ca H+K, whose rest wavelengths are shown by the dotted vertical lines).  Middle: SDSS spectra illustrating some of the AGN-fueled emission lines, plotted in the restframe of the galaxy's stars.  The wavelengths of  \mbox{H$\beta$}, [O III] $\lambda$4959, and [O III] $\lambda$5007, in the restframe of the galaxy's stars, are shown as dotted vertical lines, and the velocity shifts $\Delta v$ in the emission lines relative to the galaxy restframe are given.  Right: $50^{\prime\prime} \times 50^{\prime\prime}$ SDSS $gri$ color-composite images of the galaxies, with $5^{\prime\prime}$ scale bars shown.}}
\label{fig:examples}
\end{figure*}

Finally, we remove the AGNs with known double-peaked emission lines, which are produced by rotating gas disks, biconical outflows, or dual AGNs (e.g., \citealt{VE01.1,WH04.1,CO09.1,CR09.1,XU09.1,RO10.1,FI11.1,GR11.2}).  After removing the objects that were identified as double-peaked AGNs in searches through SDSS spectra \citep{WA09.1,LI10.1,SM10.1,GE12.1}, the result is 351 AGNs.  We measure the weighted velocity offset for each AGN as the mean of the Balmer and forbidden line-of-sight velocity offsets, weighted by their inverse variances, and the weighted velocity offsets are shown in Table~\ref{tbl-1} and Figure~\ref{fig:vhist}.

These 351 AGNs are our offset AGN candidates, and examples of the offset AGN candidates are shown in Figure~\ref{fig:examples}.  We focus the rest of the paper on analysis of the offset AGN candidates and comparisons to related groups of AGNs. 

\section{Fraction of AGNs that \\ Are Offset AGNs}
\label{small}

Since we are sensitive to only the projection of an AGN's velocity along the line of sight, our selection of offset AGN candidates is incomplete.  For instance, our criterion that offset AGN candidates have velocity offsets that are greater than 3$\sigma$ in significance excludes AGNs that have small, but real, projected velocity offsets.  Here, we estimate the fraction of all AGNs that are offset AGNs.  

First, we assume that a fraction $f_\mathrm{offset}$ of all AGNs in our parent sample are in fact offset AGNs and that every active galaxy is equally likely to host an offset AGN.  Then, we assume that an offset AGN is orbiting in the potential of the host galaxy and that the three-dimensional velocity of an offset AGN is given by the three-dimensional host galaxy velocity dispersion.  This is a reasonable assumption, since the velocity dispersion can be used as a proxy for a galaxy's gravitational potential, via the Jeans equations \citep{JE15.1}.  In this case, the observed line-of-sight velocity is $v_{obs}= \sigma_* \cos\theta$, where $\sigma_*$ is the line-of-sight velocity dispersion and $\theta$ is the polar angle of the observer to the AGN in a spherical coordinate system.  Next, we assume that the AGN's velocity has a random orientation in the plane of the host galaxy, so that $|\cos\theta|$ has a random value between 0 and 1.

\begin{figure}
\begin{center}
\includegraphics[width=8.5cm]{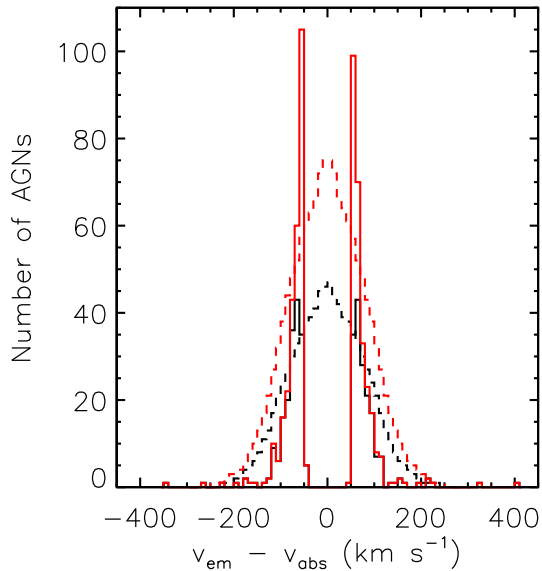}
\end{center}
\caption{Histograms of the velocity offsets of the offset AGN candidates (solid histograms) and expectations of the velocity offset distributions drawn from the AGN parent sample (dashed histograms).  The 351 offset AGN candidates are shown in black, while the red histogram includes the addition of 196 AGNs with $|v_{em}-v_{abs}|>50$ km s$^{-1}$ and that only missed offset candidate classification because their velocity errors are too large.  The black (red) dashed histograms show the expected velocity offset distributions drawn from Monte Carlo realizations, assuming that 5\% (8\%) of the parent AGN sample are offset AGNs that have random orientations in the host galaxy planes.}
\label{fig:lowv}
\end{figure}

Then, we iterate through values of $f_\mathrm{offset}$ and compare the predicted distribution of velocity offsets to the distribution of actual observed velocity offsets for the offset AGN candidates.  For each value of $f_\mathrm{offset}$, we use Monte Carlo realizations to draw 1000 distributions of predicted observed velocity offsets.  For each predicted distribution, we measure the chi-squared difference between the predicted and observed velocity offsets.  We compare the velocity offsets that have absolute values greater than 50 km s$^{-1}$. We then take the median value of the 1000 chi-squared measurements at each $f_\mathrm{offset}$ value, and  we find that $f_\mathrm{offset}=0.05$ is the best fit to the observed velocity offsets (Figure~\ref{fig:lowv}).

We also extend this approach to include AGNs that would have been classified as offset AGN candidates with $|v_{em}-v_{abs}|>50$ km s$^{-1}$, except that their line-of-sight velocity offsets are not greater than 3$\sigma$ in significance.  This adds 196 AGNs and provides an upper limit to the number of offset AGN candidates with $|v_{em}-v_{abs}|>50$ km s$^{-1}$ in our sample.  When we vary $f_\mathrm{offset}$ and match to the data, as described above, we find that $f_\mathrm{offset}=0.08$ provides the best fit to this sample (Figure~\ref{fig:lowv}).

As a test of the robustness of these figures, we also determine $f_\mathrm{offset}$ using the subsample of offset AGN candidates with velocity offsets that have absolute values greater than 70 km s$^{-1}$.  For this sample of 189 offset AGN candidates, we find $f_\mathrm{offset}=0.04$.  When we add the 18 AGNs in this velocity range that only miss offset AGN classification because their line-of-sight velocity offsets are not greater than 3$\sigma$ in significance, we find the same result.  We use this as a lower limit on the value of $f_\mathrm{offset}$.

Consequently, we estimate that 4\% -- 8\% of AGNs in our sample could in fact be offset AGNs.  With these estimates we find that our selection technique, and specifically the requirement that the velocity offsets are greater than 3$\sigma$ in significance, could be missing $\sim300 $ -- 700 offset AGNs with offset velocities $|v_{em}-v_{abs}|<50$ km s$^{-1}$.

\section{Nature of the Offset AGN Candidates}
\label{nature}

We shaped our selection criteria to select the best candidates for offset AGNs and to avoid offset emission lines produced by recoiling SMBHs or by rotating gas disks or AGN outflows, which may also conspire with dust obscuration.  Here we examine in detail whether the offset AGN candidates are consistent with being produced by recoiling SMBHs, rotating disks, AGN outflows, or dust obscuration.

\subsection{Recoiling SMBHs}

Velocity offset emission lines can also be produced by a recoiling SMBH, where the recoil is the result of gravitational wave emission after the merger of two SMBHs at the center of a merger-remnant galaxy.  However, such recoiling SMBHs, as well as subparsec-scale binary SMBHs, produce velocity offsets in the {\it broad} AGN emission lines and not the narrow AGN emission lines (e.g., \citealt{GA84.1,BO07.1,ER12.1,JU13.1,SH13.1}).  Via our criterion that the velocity offsets of the forbidden lines and the Balmer lines agree to within $1\sigma$, we have removed velocity offsets that appear in the broad lines only. 

\begin{figure}
\begin{center}
\includegraphics[width=8.5cm]{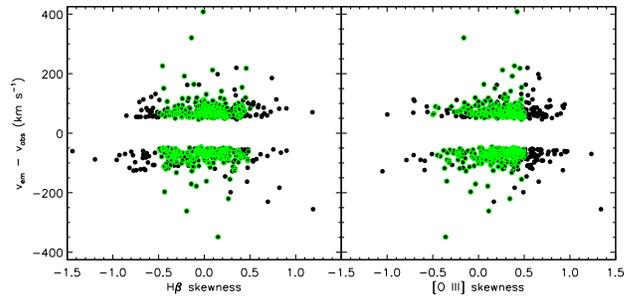}
\end{center}
\caption{The line-of-sight velocity offsets plotted against the weighted skewnesses of the \hb (left) and \oiiiw (right) emission lines.  The black points represent the velocity-shifted sample of 544 AGNs and the points circled in green represent the 351 offset AGN candidates, which are the subset of the 544 AGNs that were selected for their line symmetries and their lack of double peaks.  Negative (positive) velocity offset signals a blueshifted (redshifted) emission line, while negative (positive) skewness indicates that a line is skewed to the blue (red).  In both \hb and \oiiiwn, there are only mild correlations between blueshifted (redshifted) velocity offsets and blue (red) skewnesses. }
\label{fig:skew}
\end{figure}

\newpage 
\subsection{Rotating Disks}

Some of the velocity-offset emission lines in our sample may be caused by rotating disks, where the disk gas may be clumpy or partially obscured.  If we see only one side of the disk, then we may expect a correlation between the velocity offsets and the line asymmetries.  That is, we might expect the line to be asymmetric in the same direction of the velocity offset, with redshifted velocity offsets correlated to positive skewness and blueshifted velocity offsets correlated to negative skewness. For example, if we observe only the blueshifted side of the disk then we would observe an overall blueshift in the emission lines relative to systemic and more blue light relative to red light, which produces a negative skewness in the observed emission lines.

To test the significance of such rotating disks as a contaminant in our sample, we compare the 351 offset AGN candidates to the velocity-shifted sample of 544 AGNs (Figure~\ref{fig:skew}).  We use the skewnesses as measured in Section~\ref{selection}.  For \hbn, we find mild correlations between $v_{em}-v_{abs}$ and \hb skewness for the velocity-shifted AGN sample (the Spearman's rank correlation coefficient is 0.13 and the significance level of its deviation from zero is 0.002) and for the offset AGN candidates (Spearman's rank correlation coefficient 0.11, significance level 0.05).  For \oiiiwn, we also find mild correlations between $v_{em}-v_{abs}$ and \oiiiw skewness for the velocity-shifted AGN sample (Spearman's rank correlation coefficient 0.06, significance level 0.14) and for the offset AGN candidates (Spearman's rank correlation coefficient 0.12, significance level 0.03).

Since we find no significant correlation of velocity offset with skewness, we conclude that our sample of offset AGN candidates likely is not contaminated by many rotating disks.

\subsection{AGN Outflows}

Comparisons of the velocity offset of the narrow AGN emission line to the velocity dispersion of the host galaxy can also shed light on the source of the velocity offset.  The motion of offset AGNs is dominated by dynamical friction from the host galaxy stars, so offset AGNs should follow the potential of the host galaxy.  Due to projection effects, the measured line-of-sight velocity offset of an offset AGN should be less than or equal to the velocity dispersion.  In contrast, the measured line-of-sight velocity offset of an AGN outflow can be less than, equal to, or greater than the velocity dispersion (e.g., \citealt{NE08.1,CR10.1,TO10.1,LI13.2}).

\begin{figure}
\begin{center}
\includegraphics[width=8.5cm]{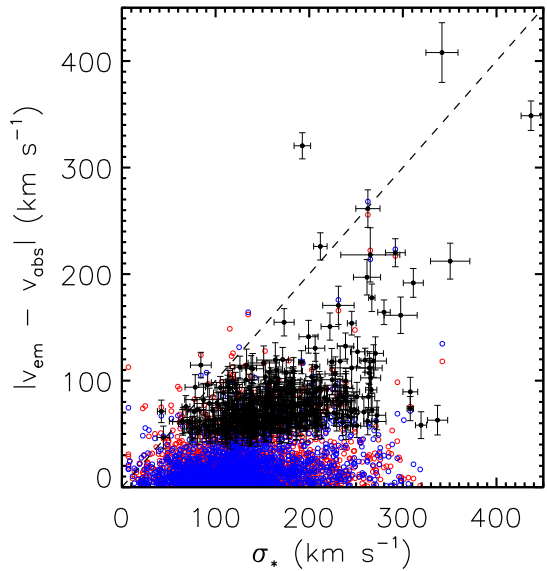}
\end{center}
\caption{Absolute values of line-of-sight velocity offsets plotted against stellar velocity dispersions.  For clarity, we show a random subsample of 10\% of the parent sample of 18314 AGNs, where the velocity offsets of the forbidden emission lines are shown in red and the velocity offsets of the Balmer emission lines are shown in blue. The black points illustrate the weighted velocity offsets of the 351 offset AGN candidates. The dashed line shows a 1:1 correlation between line-of-sight velocity offset and stellar velocity dispersion.}
\label{fig:vdisp}
\end{figure}

To learn about the nature of the SDSS offset AGN candidates, we compare the absolute values of the velocity offsets to the velocity dispersions in Figure~\ref{fig:vdisp}.  For the parent population of 18314 AGNs, 98\% (99\%) of the velocity offsets of the forbidden (Balmer) emission lines fall below the velocity dispersions.  For the 351 offset AGN candidates, 97\% of the weighted velocity offsets fall below the velocity dispersions.  

There are six offset AGN candidates with weighted velocity offsets that are $>1\sigma$ above the velocity dispersions.  These may be turbulent merging galaxies where the velocity dispersion is changing rapidly (e.g., \citealt{JO09.1,ST12.1}), or they may be examples of AGN outflow interlopers.  Follow-up observations are necessary to determine their true natures (Section~\ref{conclusions}).   

The relative numbers of redshifted and blueshifted AGN emission lines also offer clues into the sources of the velocity offsets.  If offset AGNs have random velocity projections to the line of sight, then there should be equal numbers of offset AGNs with observed redshifts and offset AGNs with observed blueshifts.  In contrast, AGN outflows are known to preferentially result in observations of blueshifted emission lines, since redshifted lines are often obscured by the AGN torus (e.g., \citealt{ZA02.1}).  Among our offset AGN candidates, 170 ($48^{+3}_{-2}\%$) exhibit redshifted lines and 181 ($52^{+2}_{-3}\%$) exhibit blueshifted lines, which is consistent with the equal numbers of redshifted and blueshifted observations expected for offset AGNs.

\begin{figure*}
\begin{center}
\includegraphics[width=18cm]{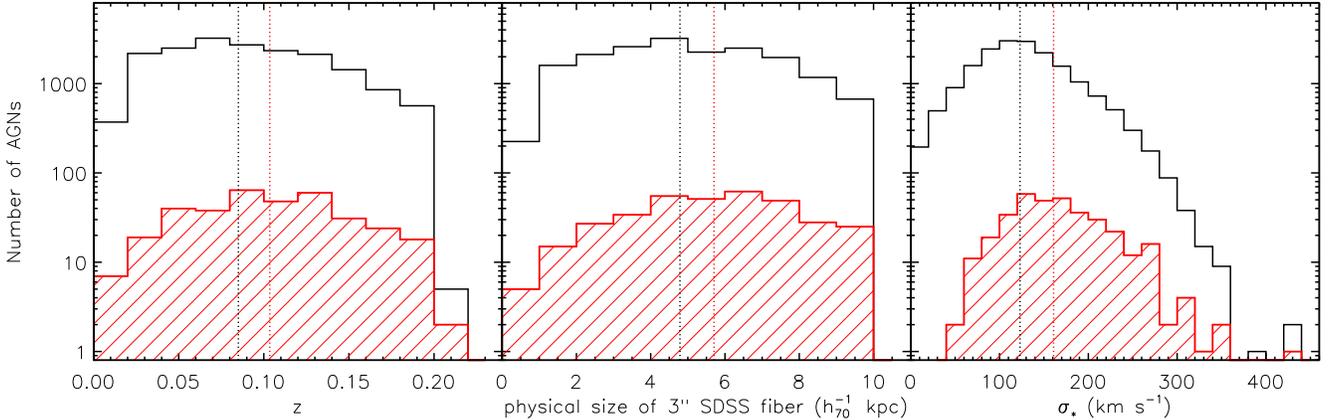}
\end{center}
\caption{Histograms of the distributions of the parent sample of 18314 AGNs (black) and the 351 offset AGN candidates (red) in redshift (left), physical size of the SDSS fiber (middle), and velocity dispersion (right).  For each distribution the median value is marked with a dotted vertical line.}
\label{fig:z_kpc_vdisp}
\end{figure*}

\subsection{Dust Obscuration}

If dust obscures one of the AGN in a dual AGN system, this could lead us to detect an offset AGN.  Or, dust could conspire with disk rotation or AGN outflows to create the velocity offsets that we measure, by obscuring only the redshifted or blueshifted component of emission from a disk or outflow.  We note that dust on small scales around the AGN torus should not affect our sample, which consists of Type 2 AGNs.  In general the amount of larger scale dust obscuration increases with a galaxy's inclination, with the most obscuration occurring for edge-on galaxies, which have inclinations of 90$^\circ$.  

As a test of whether dust is a large contaminant in our offset AGN candidates, we measure the inclinations of their host galaxies and compare to the inclinations of the host galaxies of the AGN parent population. We measure a galaxy's inclination via the exponential fit to the ratio of semiminor to semimajor axis $b/a$ in $r$ band, given by {\tt expAB\_r} with error {\tt expABErr\_r}, in the SDSS PhotObj catalog.  The median error on the inclination for the AGN sample is $0.7^\circ$.  We find that the offset AGN candidates have typical inclinations (mean inclination $51^\circ$, with a standard deviation of $14^\circ$) that are consistent with those of the AGN parent sample (mean inclination $48^\circ$, with a standard deviation of $15^\circ$).  

We also test the dust obscuration hypothesis by examining the AGN luminosities as a function of host galaxy inclinations.  If dust plays a significant role in our offset AGN sample, then we expect more obscured (fainter) AGNs as inclination increases.  Instead, we find that the AGN bolometric luminosities are independent of inclination (Spearman's rank correlation coefficient 0.06, significance level 0.26).

Based on the distributions of inclinations and AGN luminosities, we find no evidence for dust obscuration as a large contributor to the velocity offsets we measure in the offset AGN candidates.
As a more thorough test of the role of dust obscuration in producing velocity-offset AGN emission features, we suggest that a large, complete sample of integral field spectrograph observations of active galaxies would be useful.  Artificial dust reddening could be added to the data in different spatial locations on the galaxy in a systematic fashion, as a test of the effect on the observed stellar absorption and AGN emission kinematics.

\section{Results}

\subsection{Comparison to Parent AGN Population}

The offset AGN candidates reflect the parent AGN population in overall distributions of redshift, physical size of the SDSS fiber, and stellar velocity dispersion, but the median values differ (Figure~\ref{fig:z_kpc_vdisp}).
The median stellar velocity dispersion in the offset AGN candidates' host galaxies ($\sigma_*=161$ km s$^{-1}$) is higher than the median stellar velocity dispersion of the parent population ($\sigma_*=123$ km s$^{-1}$).  This could be explained if the offset AGN sample consists of galaxy mergers that exhibit high velocity dispersions because the stars are not yet dynamically relaxed.  Or, the higher stellar velocity dispersions  could reflect a selection effect from our approach to identifying offset AGN candidates.  We select offset AGN candidates as AGNs that have velocity offsets that are greater than $3\sigma$ in significance, which eliminates AGNs with velocity offsets $\lesssim 50$ km s$^{-1}$.  If we assume that offset AGNs follow the orbital velocity of the host galaxy stars and that the offset AGNs' orbits are randomly oriented in the host galaxy planes (Section~\ref{small}), then our sample of offset AGN candidates is biased towards galaxies with higher velocity dispersions.  

We also find that the offset AGN candidates have somewhat higher redshifts (median $z=0.104$) than the parent AGN population (median $z=0.085$) and, correspondingly, the physical size of the SDSS fiber is somewhat larger on average (a median of 5.7 h$_{70}^{-1}$ kpc for the offset candidates, compared to a median of 4.8 h$_{70}^{-1}$ kpc for the parent population).  This may reflect the bias towards higher $\sigma_*$ for the offset AGN candidates, as discussed above, if $\sigma_*$ correlates with redshift in the SDSS catalog (e.g., \citealt{TH13.1}). Further, if the offset AGN catalog contains a higher fraction of galaxy mergers, the higher median redshift may reflect the increasing incidence of galaxy mergers with redshift (e.g., \citealt{CO03.2,LI08.1,LO11.1}).

\begin{figure*}
\begin{center}
\includegraphics[width=18cm]{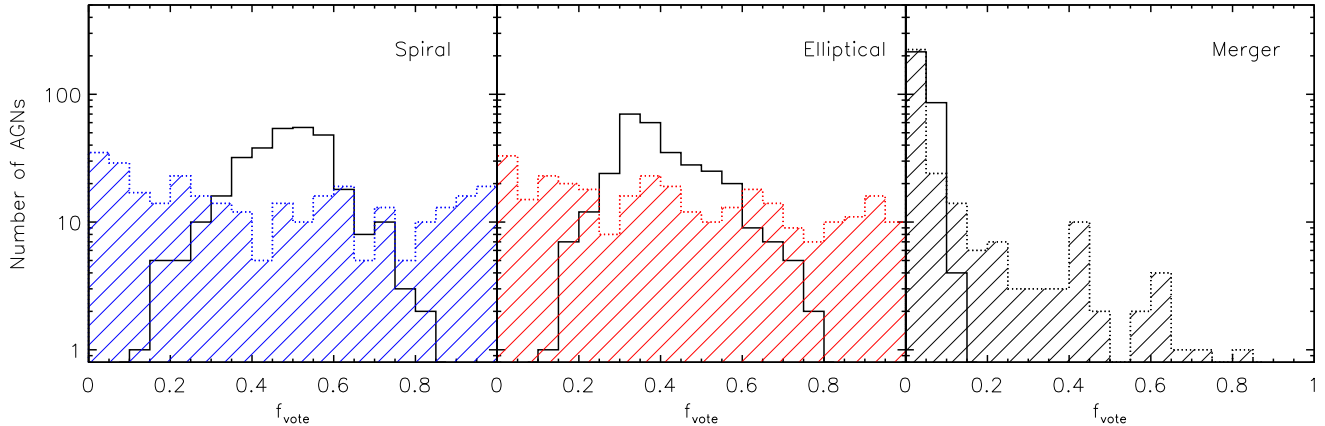}
\end{center}
\caption{Histograms of ${f}_\mathrm{vote}$, the fraction of Galaxy Zoo votes for a  morphology classification for the offset AGN candidates (dotted histograms) and the mean fraction of votes for the same morphology classification for a comparison sample of AGNs matched in color, stellar mass, and redshift (solid histograms).  The vote fractions for spiral, elliptical, and merger classifications are shown.}
\label{fig:gz}
\end{figure*}

\subsection{Morphologies and Environment}

If the offset AGN candidates are in fact offset AGNs in ongoing galaxy mergers, then we expect the candidates to preferentially reside in host galaxies that have merger morphologies or elliptical morphologies (since merger remnants have properties similar to ellipticals; e.g., \citealt{BA88.2,HE92.1}) and reside in dense environments with many other galaxies.  Color and stellar mass are well known to correlate with morphology and environment.
Massive, red galaxies preferentially occur with elliptical morphologies, while less massive, blue galaxies preferentially occur with spiral morphologies (e.g., \citealt{DE61.1,ST01.2}).   Furthermore, red galaxies are more clustered with other galaxies, while blue galaxies are less clustered with other galaxies (e.g., \citealt{BO92.2,YA05.2,CO06.3}).  Finally, the occurrence rate of galaxy mergers has been observed to increase with redshift (e.g., \citealt{CO03.2,LI08.1,LO11.1}).  We account for these underlying trends in morphology by comparing the offset AGN candidates to samples of SDSS AGNs that are matched in color, stellar mass, and redshift.  Consequently, any remaining trends with morphology or environment can be attributed to the offset AGN candidates themselves.

For each offset AGN candidate, we build a comparison sample of AGNs (selected from the SDSS DR7 catalog of $z<0.21$ galaxies, as described in Section~\ref{selection}) that have the same $u-r$ color, stellar mass, and redshift, to within $10\%$, as the candidate.  We construct these comparison samples for the 349 of the 351 offset AGN candidates that have existing stellar mass measurements based on fits to the photometry \citep{KA03.2,SA07.2}, which we obtain from the MPA-JHU SDSS DR7 data product\footnote{\url{http://www.mpa-garching.mpg.de/SDSS/}}.  Colors are obtained from $u$ and $r$ magnitudes measured in the SDSS photometry pipeline. We choose the $10\%$ range because it produces comparison sample sizes of $\sim100$ for most offset AGN candidates; the median number of AGNs in a comparison sample is 89.  There are 44 offset AGN candidates that have comparison groups with fewer than 10 members, and we do not include them in the morphology analysis.

Here, we examine the morphologies and environments of the remaining 305 offset AGN candidates that have comparison samples of at least 10 AGNs.  We find a small overall preference (with large errors) for offset AGN candidates to occur in galaxies with elliptical, S0, and merger morphologies, as well as a bias for offset AGN candidates, on average, to occur in more clustered environments.

\begin{figure}
\begin{center}
\includegraphics[width=8.5cm]{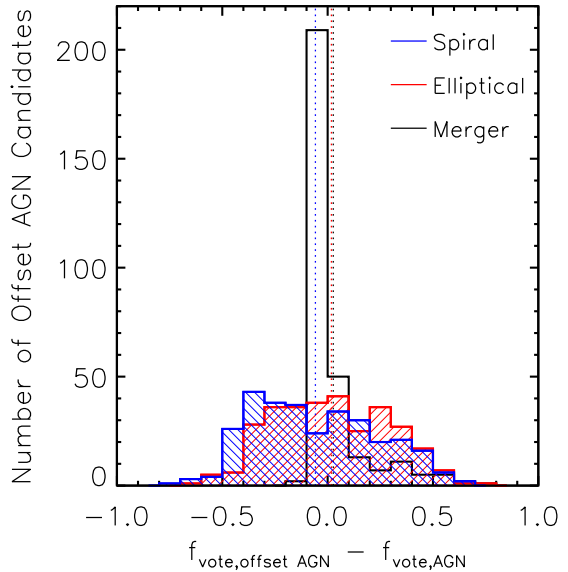}
\end{center}
\caption{Histograms of $f_\mathrm{vote,offset \, AGN}-\bar{f}_\mathrm{vote,AGN}$, the differences between the fraction of Galaxy Zoo votes for a  morphology classification for an offset AGN candidate $f_\mathrm{vote,offset \, AGN}$ and the mean fraction of Galaxy Zoo votes for the same morphology classification for a comparison sample of AGNs matched in color, stellar mass, and redshift $\bar{f}_\mathrm{vote,AGN}$.  Fractions of the Galaxy Zoo votes for spiral (blue), elliptical (red), and merger (black) classifications are shown.  For each distribution the mean value is marked with a dotted vertical line.}
\label{fig:gz_differences}
\end{figure}

\subsubsection{Morphology}

We examine the host galaxy morphologies of the offset AGN candidates using two different visual morphology indicators: one that is qualitative, and one that is more quantitative.  The qualitative approach to visual morphologies is Galaxy Zoo, a catalog of visual morphology classifications assigned by citizen scientists to $\sim900,000$ galaxies in SDSS DR6 \citep{LI11.1}.  Multiple citizen scientists vote to classify each galaxy as an elliptical, clockwise spiral, anti-clockwise spiral, edge on spiral galaxy, or merger.  A known bias is that galaxies that are distant, small, or faint are more likely to be classified as ellipticals, even though they may be spiral galaxies with arms that are difficult to discern \citep{BA09.1}.  A bias correction has been applied to account for this effect, based on the assumption that the fractions of galaxies with each morphology classification remains constant for a given galaxy size and luminosity.  This correction resulted in a catalog of debiased vote fractions for the galaxies \citep{LI11.1}.  We use the debiased vote fractions of elliptical, combined spiral (which combines the three types of spiral classifications), and merger for our analysis.

\begin{figure}
\begin{center}
\includegraphics[width=8.5cm]{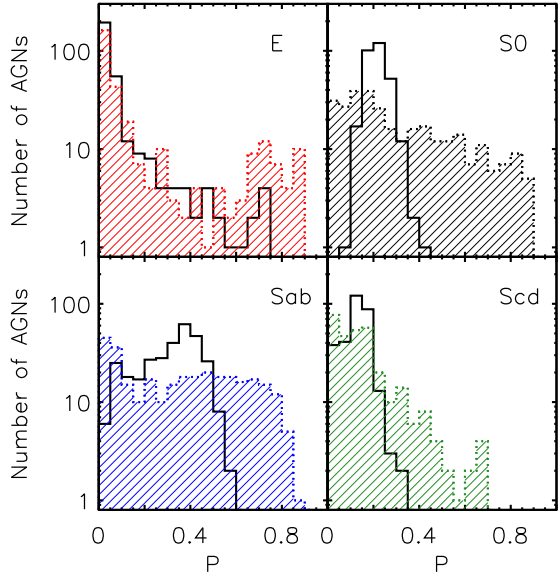}
\end{center}
\caption{Histograms of the probabilities $P$ of a Bayesian automated morphology classification for the offset AGN candidates (dotted histograms) and the median probabilities of the same morphology classification for a comparison sample of AGNs matched in color, stellar mass, and redshift (solid histograms).  The probabilities for elliptical, S0, Sab, and Scd classifications are shown.}
\label{fig:bayesian}
\end{figure}

For each morphology category, we compare the offset AGN candidate's vote fraction to the mean vote fraction of the same classification for the AGN comparison sample (Figure~\ref{fig:gz}). On average, the offset AGN candidates received a slightly higher fraction (with large errors) of votes for elliptical and merger morphologies and a lower fraction of votes for spiral morphologies (Figure~\ref{fig:gz_differences}).  On average, the offset AGN candidates have $2 \pm 27\%$ ($3 \pm 14\%$) higher vote fractions for elliptical (merger) morphologies than their comparison AGN samples.  Conversely, the offset AGN candidates have $6 \pm 29\%$ lower vote fractions, on average, for spiral morphologies than their comparison AGN samples.

We also assess morphology using a catalog of Bayesian automated morphology classifications for $\sim700,000$ galaxies in SDSS DR7 \citep{HU11.1}, and this catalog takes a more quantitative approach to morphology classifications.  This technique, which uses a machine learning algorithm trained on visual classifications of morphologies, assigns each galaxy a probability of being an elliptical (E), S0, early-type spiral (Sab), or late-type spiral (Scd) galaxy.

\begin{figure}
\begin{center}
\includegraphics[width=8.5cm]{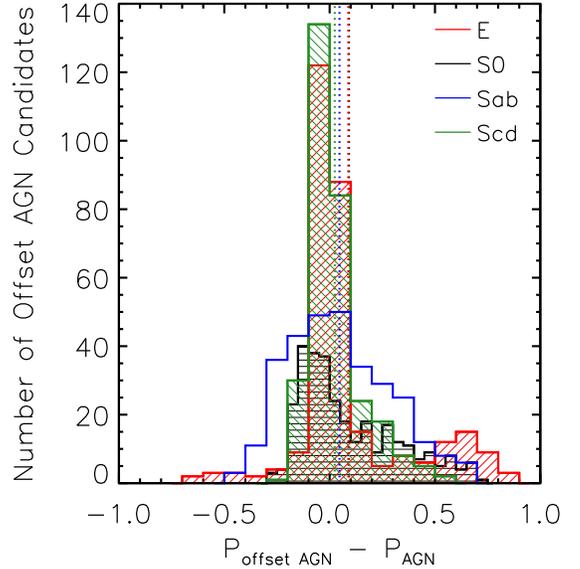}
\end{center}
\caption{Histograms of $P_\mathrm{offset \, AGN}-\tilde{P}_\mathrm{AGN}$, the differences between the probabilities of a Bayesian automated morphology classification for an offset AGN candidate $P_\mathrm{offset \, AGN}$ and the median probability of the same morphology classification for a comparison sample of AGNs matched in color, stellar mass, and redshift $\tilde{P}_\mathrm{AGN}$.  Probabilities for elliptical (red), S0 (black), Sab (blue), and Scd (green) classifications are shown.  For each distribution the mean value is marked with a dotted vertical line.}
\label{fig:bayesian_difference}
\end{figure}

For each morphology class, we compare the offset AGN candidate's probability to the median probability of the same classification for the AGN comparison sample (Figure~\ref{fig:bayesian}).  We find that, on average, the offset AGN candidates have ($9 \pm 26\%$, $9 \pm 23\%$, $5 \pm 24\%$, $2 \pm 13\%$) higher probabilities of (E, S0, Sab, Scd) classifications (Figure~\ref{fig:bayesian_difference}).  The spread in probabilities for each morphology category is large, but overall the offset AGN candidates have the greatest enhancement in probabilities for the elliptical and S0 classifications.    

In summary, the two morphology catalogs we use yield similar results.  We find enhanced probabilities of (2--9\%,9\%,3\%), with large error bars, for an offset AGN candidate to reside in a galaxy with an elliptical, S0, or merger morphology, respectively, when compared to AGNs with similar colors, stellar masses, and redshifts.

\subsubsection{Environment}

To study the global environments of the offset AGN candidates, we examine the number of galaxies $N_{group}$ in each candidate's group.  We take group membership from an SDSS DR7 group catalog \citep{TI11.2,WE12.1}, which uses the halo-based group finding algorithm of \cite{YA05.1}.  This catalog identifies groups only to $z=0.1$, due to the limitations in robust group determination at higher redshifts.  

Because of the redshift limit of the group catalog, only 117 offset AGN candidates are in the group catalog with comparison samples of at least 10 AGNs matched in color, stellar mass, and redshift.  Of these 117 offset AGN candidates, 101 have $N_{group} \leq 5$ (field), 14 have $5 < N_{group} \leq 50$ (group), and 2 have $N_{group} > 50$ (cluster).  

\begin{figure}
\begin{center}
\includegraphics[width=8.5cm]{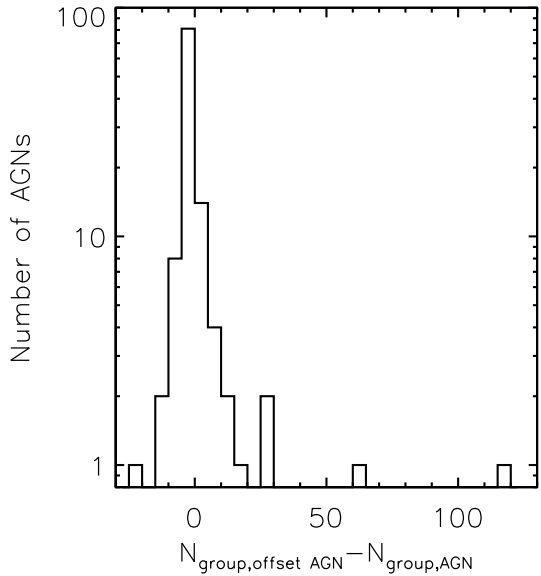}
\end{center}
\caption{Histogram of $N_\mathrm{group,offset \, AGN}-\bar{N}_\mathrm{group,AGN}$, the differences between the number of group members for an offset AGN candidate $N_\mathrm{group,offset \, AGN}$ and the mean number of group members for its comparison sample of AGNs matched in color, stellar mass, and redshift $\bar{N}_{group,AGN}$.  A positive value of $N_\mathrm{group,offset \, AGN}-\bar{N}_\mathrm{group,AGN}$ indicates that an offset AGN candidate resides in a richer environment than the average environment of its comparison sample.}
\label{fig:environment}
\end{figure}

When we examine the differences between $N_{group}$ for an offset AGN candidate and the mean $\bar{N}_{group}$ for its comparison sample (Figure~\ref{fig:environment}), we find that 104 offset AGN candidates reside in groups with the same number of members, within 1$\sigma$, as the mean.  Of the cases where $N_{group}$ for an offset AGN candidate differs by $>1\sigma$ from $\bar{N}_{group}$ for its comparison sample, there are 13 offset AGN candidates in richer environments and no offset AGN candidates in poorer environments.  This hints towards a tendency for offset AGN candidates to live in more clustered environments than other AGNs.

\begin{figure}
\begin{center}
\includegraphics[width=8.5cm]{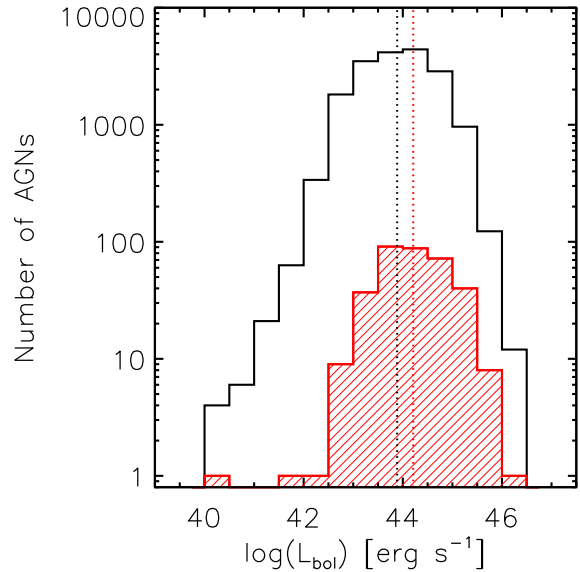}
\end{center}
\caption{Histograms of the bolometric luminosities of the AGN samples.  The parent sample of 18314 AGNs is illustrated by the solid black histogram, while the 351 offset AGN candidates are illustrated by the red histogram. For each distribution the mean value is marked with a dotted vertical line.}
\label{fig:lhist}
\end{figure}

\subsection{Fraction of Offset AGN Candidates Increases with AGN Luminosity}

AGN activity can be powered by gas inflows driven by galaxy mergers (e.g., \citealt{HE84.1,SA88.2,SP05.2}) as well as stochastic accretion of gas (e.g., \citealt{NO83.1,MA99.3,HO06.1,LI08.3}), but the relative roles of these two mechanisms and the dependence on AGN luminosity is not clear.  It has been suggested that galaxy mergers may play a more significant role in fueling higher-luminosity AGNs (e.g., \citealt{HO09.1,TR12.1}), while other studies indicate that there is no correlation of mergers with AGN luminosity (e.g., \citealt{VI14.1}). 

Based on visual classification of mergers, \cite{TR12.1} find that the fraction of AGNs associated with mergers increases with the AGN bolometric luminosity, from $\sim4\%$ at $\log(L_\mathrm{bol})\sim43$ to $\sim70\%$ at $\log(L_\mathrm{bol})\sim46$.  However, this result is based on identifying mergers via visual morphology classifications of the host galaxies.  It is difficult to build a clean sample of AGNs in mergers with this approach, given that there are known limitations in using visual morphologies to identify mergers and these limitations can lead to significant numbers of false classifications (e.g., \citealt{LO11.1}).

Offset AGNs are uniquely well suited to building a clean sample of AGNs in mergers, since offset AGNs are by definition actively accreting supermassive black holes in ongoing galaxy mergers.  We make the first attempt at using offset AGNs for studies of merger-triggered AGN activity, using our sample of offset AGN candidates.  A final robust measurement will be possible once we have used follow-up observations to identify the true offset AGNs in the sample.

First, we determine the AGN bolometric luminosities by converting the OSSY measurements of \oiiiw fluxes to \oiiiw luminosities, then using the observed scaling relation of AGN bolometric luminosity with \oiiiw luminosity \citep{HE04.3}.  Next, we compare the AGN bolometric luminosities of the offset AGN candidates to the parent AGN sample (Figure~\ref{fig:lhist}).  The mean luminosity of the offset AGN candidates is $\sim0.3$ dex higher ($\log(L_\mathrm{bol})=44.2 \pm 0.7$) than that of the parent AGNs ($\log(L_\mathrm{bol})=43.9 \pm 0.7$), although they are consistent to within 1$\sigma$.  

\begin{figure}
\begin{center}
\includegraphics[width=8.5cm]{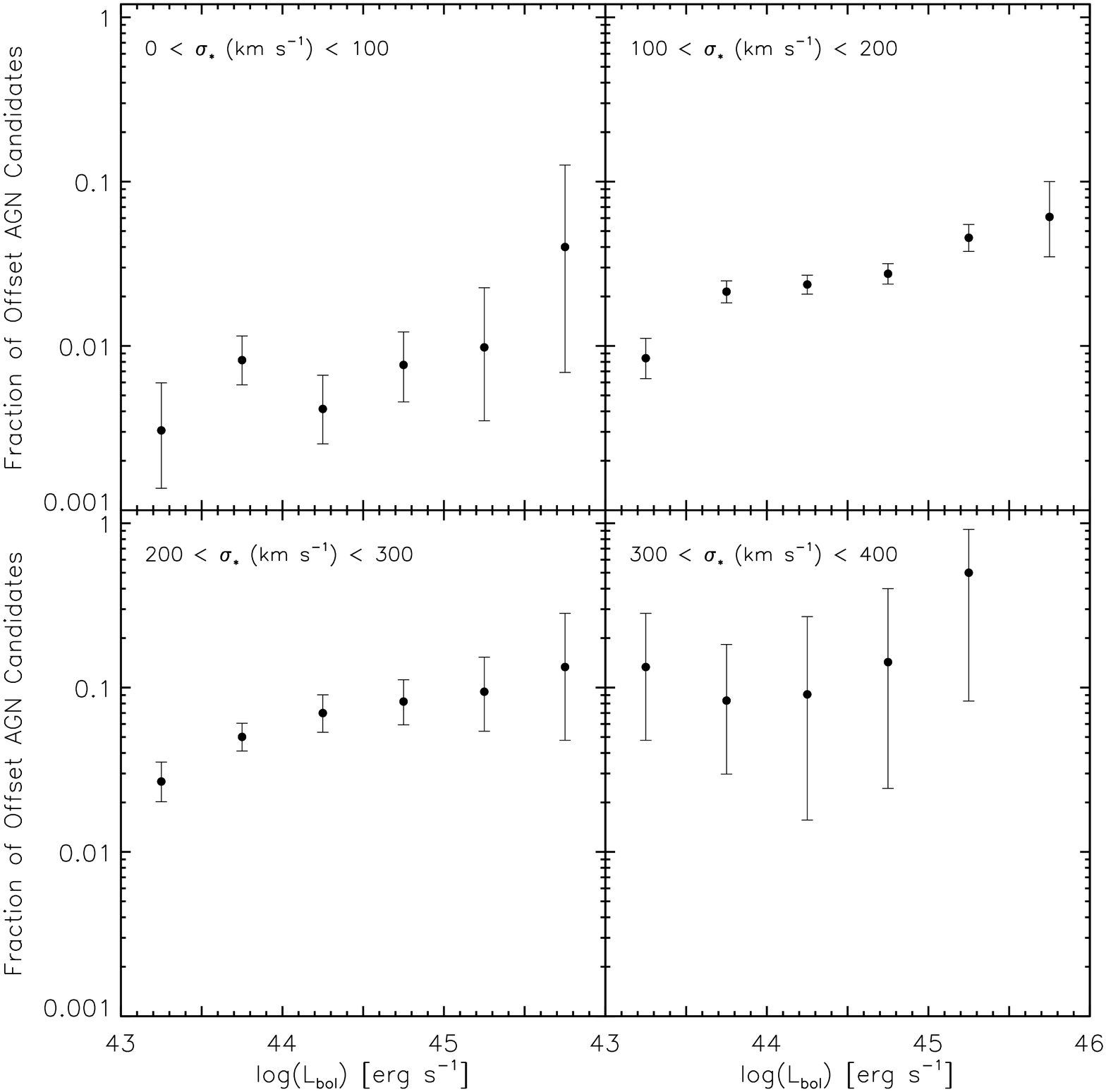}
\end{center}
\caption{The fraction of offset AGN candidates in the parent AGN sample, binned by velocity dispersion and then binned by bolometric luminosity.  In the fourth panel, there are no parent AGNs in the bin $45.5 < \log(L_\mathrm{bol})$ [erg s$^{-1}$] $< 46$. The fraction of offset AGN candidates increases with $L_\mathrm{bol}$ and with $\sigma_*$.}
\label{fig:offset_bin}
\end{figure}

Since AGN bolometric luminosity may scale weakly with velocity dispersion (e.g., \citealt{WO02.1}), we bin the AGNs by host galaxy velocity dispersion before examining how the offset AGN candidate fraction depends on bolometric luminosity.  We then bin by bolometric luminosity and find that the fraction of offset AGN candidates in the parent AGN sample increases with AGN bolometric luminosity for each range in velocity dispersion (Figure~\ref{fig:offset_bin}).  

\begin{figure}
\begin{center}
\includegraphics[width=8.5cm]{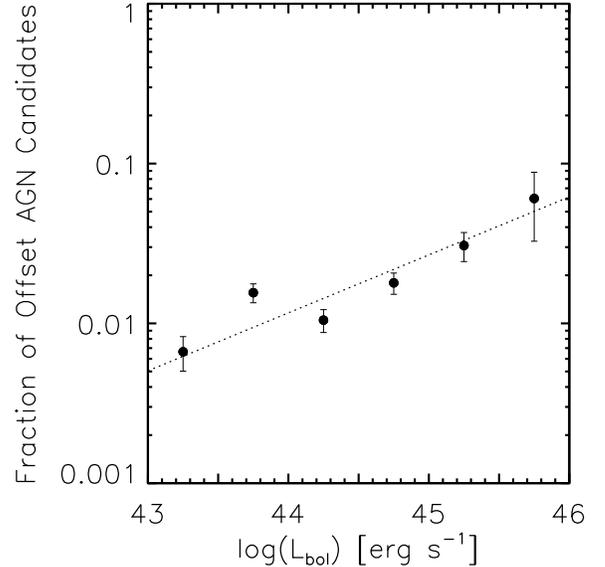}
\end{center}
\caption{The fraction of offset AGN candidates in the parent AGN sample, binned by velocity dispersion and then stacked and binned by bolometric luminosity.  The velocity dispersion binning accounts for the bias towards selecting offset AGN candidates in galaxies with higher velocity dispersions.  The dotted line shows the best-fit power law, $f=(L_\mathrm{bol}/2.1 \times 10^{49} \mathrm{\, erg \, s}^{-1})^{0.36}$.}
\label{fig:offset_stack}
\end{figure}

Finally, we stack the four velocity dispersion bins, using the mean of the offset AGN candidate fractions (in each bolometric luminosity bin) weighted by their inverse variances.  We find that the fraction of offset AGN candidates in the parent AGN sample increases by an order of magnitude with AGN bolometric luminosity, from $0.7\%$ to $6\%$ over the bolometric luminosity range $43 < \log(L_\mathrm{bol})$ [erg s$^{-1}$] $< 46$ (Figure~\ref{fig:offset_stack}).  The best-fit power law to the fraction of offset AGN candidates, $f$, is

\begin{equation}
f=\left( \frac{L_\mathrm{bol}} {2.1 \times 10^{49} \mathrm{\, erg \, s}^{-1}} \right)^{0.36} \; .
\end{equation}

The general trend of increasing fraction of AGNs in mergers with an increasing AGN bolometric luminosity is the same as seen in \cite{TR12.1} over the same bolometric luminosity range.  \cite{TR12.1} also fit a power law to their $f$ -- $L_\mathrm{bol}$ relation, and their exponent is the same ($0.4$) but their amplitude is lower ([$3 \times 10^{46}$ erg s$^{-1}]^{-0.4}$).  However, we note that the true amplitude of our relation is not well constrained due to known biases in our selection method (Section~\ref{small}). 

If these candidates are in fact offset AGNs, then the observed trend is direct observational evidence that galaxy mergers preferentially trigger high-luminosity AGNs. However, if some of the offset AGN candidates are AGN outflows, then the relation may be explained in part by outflows that are correlated with higher bolometric luminosities (e.g., \citealt{HA12.1,KI13.1,TO13.1,CI14.1}).  Forthcoming follow-up observations will confirm the bona fide offset AGNs and enable a robust measurement of any trend of increasing offset AGN fraction with AGN bolometric luminosity.

\section{Comparison to Other Samples of Offset AGN Candidates}

We now compare to higher redshift samples of offset AGN candidates that were selected in the AGES and DEEP2 surveys \citep{CO09.1,CO13.1} using similar criteria to those applied to SDSS here.  The offset AGN candidates in both AGES and DEEP2 have line-of-sight velocity offsets of forbidden and Balmer lines that agree to $1\sigma$ and that are $>3\sigma$ in significance.  In SDSS, we find 351 offset AGN candidates ($1.9^{+0.1}_{-0.1}\%$ of AGNs) with 50 km s$^{-1} < |v_{em}-v_{abs}| <$ 410 km s$^{-1}$ at a mean redshift $\bar{z}=0.1$.  In AGES, there are 5 offset AGN candidates ($2.9^{+1.9}_{-0.8}\%$ of AGNs) with 100 km s$^{-1} < |v_{em}-v_{abs}| <$ 220 km s$^{-1}$ at a mean redshift $\bar{z}=0.25$.  In DEEP2, there are 30 offset AGN candidates ($33^{+5}_{-5}\%$ of AGNs in red galaxies) with 40 km s$^{-1} < |v_{em}-v_{abs}| <$ 270 km s$^{-1}$ at a mean redshift $\bar{z}=0.7$.  

After accounting for projection effects that remove small line-of-sight velocity offsets from the offset AGN candidate samples, we find that $4\% - 8\%$ of SDSS AGNs could be offset AGNs (Section~\ref{small}) while $44\% - 60\%$ of DEEP2 AGNs could be offset AGNs \citep{CO09.1}.  One possibility for this increase is that effects such as outflows and dust obscuration, which can produce velocity-offset AGN emission lines (Section~\ref{nature}), may play a greater role in higher-redshift AGNs.

Before comparing the offset AGN candidates between these three surveys, we must account for differences in the surveys' spatial scales, spectral resolutions, and parent AGN selections.  One possible bias in the comparisons is that the surveys are probing emission produced on different spatial scales.  SDSS uses $3^{\prime\prime}$ fibers that correspond to 5.5 kpc at typical redshift $z=0.1$, AGES uses $1\farcs5$ fibers that correspond to 5.0 kpc at typical redshift $z=0.2$, and DEEP2 uses slits with 7$^{\prime\prime}$ average lengths that correspond to 50 kpc at typical redshift $z=0.6$ of the offset AGN candidates.  The dual AGN candidates in SDSS and DEEP2 have typical spatial separations of $\sim1$ kpc \citep{CO09.1,CO12.1}, so if the offset AGN candidates are related systems (dual supermassive black holes, but where only one is an AGN) then their spatial separations should be $\lesssim1$ kpc.  Consequently, the different spatial scales probed by the three surveys would not bias the relative numbers of offset AGN candidates found in each.

A more significant bias arises from the different spectral resolutions of the three surveys ($R\sim1800$ for SDSS, $R\sim1000$ for AGES, and $R\sim5000$ for DEEP2).  The high spectral resolution of DEEP2 enables resolution of much smaller velocity offsets than is possible with SDSS or AGES.  Further, the DEEP2 search for offset AGN candidates was limited to AGNs in red host galaxies.  In order to compare offset AGN candidates across the redshift range of the SDSS and DEEP2 surveys, we select the offset AGN subsamples with velocity offsets $|v_{em}-v_{abs}|>46$ km s$^{-1}$ (the minimum velocity offset detected by SDSS, which has a lower spectral resolution than DEEP2) and that reside in host galaxies matched in color ($-0.6 < u-r < 5.8$) and rest-frame absolute magnitude ($-23.0 < \mathrm{M}_r < -18.3$) to the DEEP2 AGN sample.

When we match on minimum velocity offset, color, and rest-frame absolute magnitude in this way, we find that $1.9^{+0.1}_{-0.1}\%$ (345/18105) of AGNs are offset AGN candidates in SDSS and $32^{+6}_{-5}\%$ (29/91) of AGNs are offset AGN candidates in DEEP2.   Hence, we find that the fraction of offset AGN candidates increases by a factor of $\sim16$ from $z=0.1$ to $z=0.7$.  This trend can be understood if these candidates are indeed offset AGNs produced by galaxy mergers, since the galaxy merger fraction also increases with redshift (e.g., \citealt{CO03.2,LI08.1,LO11.1}).

\section{Conclusions}
\label{conclusions}

We have searched SDSS for an expected kinematic signature of offset AGNs: galaxy spectra that have AGN emission lines that are offset in line-of-sight velocity from the stellar absorption lines.  Offset AGNs are produced when a merger between two galaxies brings two supermassive black holes into a merger remnant galaxy, and one of the supermassive black hole is fueled as an AGN.  We find 351 offset AGN candidates out of a sample of 18314 AGNs at $z<0.21$, and their velocity offsets span 50 km s$^{-1} < |v_{em}-v_{abs}| <$ 410 km s$^{-1}$.  We examine the characteristics of these offset AGN candidates and what they reveal about galaxy mergers, and our main results are summarized below.

1. By assuming that the AGN velocity offsets we measure reflect the host galaxy stellar velocity dispersions and are randomly oriented in the host galaxy planes, we estimate that $4\%$ -- $8\%$ of AGNs are offset AGNs.

 2.  We find that the offset AGN candidates, when compared to AGNs matched in color, stellar mass, and redshift, are more likely on average (with large error bars) to reside in host galaxies with elliptical or merger morphologies.  We also uncover hints that offset AGN candidates reside in more clustered environments than the AGNs matched in color, stellar mass, and redshift.  These host galaxy morphologies and denser environments are consistent with expectations for offset AGNs, which are produced by mergers.
 
 3.  The fraction of AGNs that are offset AGN candidates increases with AGN bolometric luminosity, from $0.7\%$ to $6\%$ over the luminosity range $43 < \log(L_\mathrm{bol})$ [erg s$^{-1}$] $< 46$.  This suggests that higher luminosity AGNs are preferentially triggered by galaxy mergers and not stochastic accretion processes, and a robust measurement of this trend will be possible with future confirmations of offset AGNs.
 
 4. By comparing and carefully matching to offset AGN candidates in galaxy surveys at higher redshifts, we find that the fraction of AGNs that are identified as offset AGN candidates increases from $1.9^{+0.1}_{-0.1}\%$ at $z=0.1$ to $32^{+6}_{-5}\%$ at $z=0.7$, a factor of $\sim16$ increase over that redshift range.  This trend is suggestive of offset AGNs that are produced by galaxy mergers and so trace the galaxy merger rate, which exhibits similar increases with redshift over this range.
 
We developed our offset AGN candidate selection criteria carefully to select for velocity offsets produced by offset AGNs and against velocity offsets produced by gravitational recoil of a SMBH, an outflow from a single AGN, or a rotating gas disk around a single AGN.  Multiple tests indicate that the candidates have the traits of offset AGNs and not of recoiling SMBHs, outflows, or disk rotation, but follow-up observations are required to confirm the true nature of the candidates.  Spatial resolution of AGNs that are spatially offset from their host galaxy centers would provide direct confirmation of offset AGNs, and could be accomplished through spatially resolved spectroscopy, X-ray observations, or radio observations to pinpoint the spatial locations of the AGNs.  
 
The selection of offset AGN candidates represents the first step towards building a catalog of confirmed offset AGNs, which would be the first of its kind and usher in a new angle of approach to observational studies of galaxy mergers and AGN fueling.  Such studies are often limited by the difficulty of observationally identifying AGNs associated with ongoing galaxy mergers. A catalog of offset AGNs, built from candidates like those presented here, will enable unparalleled precision in resolving the link between galaxy mergers and AGN activity because offset AGNs are, by definition, actively accreting supermassive black holes in the midst of galaxy mergers. 
 
\acknowledgements{We thank the referee for a prompt and useful report.  We also gratefully acknowledge insightful discussions with Scott Barrows, Karl Gebhardt, Francisco M\"uller S\'anchez, and Kyuseok Oh.} 

\newpage

\bibliographystyle{apj}

\end{document}